# An introduction to nanothermodynamics:
# Thermal equilibrium for heterogeneous and finite-sized systems

Ralph V. Chamberlin, Michael R. Clark
Department of Physics, Arizona State University, Tempe, AZ  85287-1504 USA
Vladimiro Mujica, and George H. Wolf
School of Molecular Science, Arizona State University, Tempe, AZ  85287-1604 USA

**Abstract**

The theory of small-system thermodynamics was originally developed to extend the laws of thermodynamics to length scales of nanometers. Here we review this "nanothermodynamics," and stress how it also applies to large systems that subdivide into a heterogeneous distribution of internal subsystems that we call "regions." We emphasize that the true thermal equilibrium of most systems often requires that these regions are in the fully-open generalized ensemble, with a distribution of region sizes that is not externally constrained, which we call the "nanocanonical" ensemble. We focus on how nanothermodynamics impacts the statistical mechanics of specific models. One example is an ideal gas of indistinguishable atoms in a large volume that subdivides into an ensemble of small regions of variable volume, with separate regions containing atoms that are distinguishable from those in other regions. Combining such subdivided regions yields the correct entropy of mixing, avoiding Gibbs paradox without resorting to macroscopic quantum symmetry for semi-classical particles. Other models are based on Ising-like spins (binary degrees of freedom), which are solved analytically in one-dimension, making them suitable examples for introductory courses in statistical physics. A key result is to quantify the net increase in entropy when large systems subdivide into small regions of variable size. Another result is to show similarity in the equilibrium properties of a two-state model in the nanocanonical ensemble and a three-state model in the canonical ensemble. Thus, emergent phenomena may alter the thermal behavior of microscopic models, and the correct ensemble is necessary for accurate predictions.

## I. Introduction

Thermodynamics is arguably the broadest theory in all of science [1]. Indeed, the two main laws of thermodynamics: 1) that total energy is conserved and 2) that total entropy does not decrease, should apply to quantum and classical systems on size scales from subatomic particles to the entire universe. In addition, thermodynamics should agree with all experiments, and other methods of scientific research such as theory from statistical mechanics and simulations from computers. However, over the past few decades some inconsistencies have been found between experiments, theory, and simulations. For example, measurements related to non-resonant spectral hole burning show that excess energy added to most types of materials – including liquids, glasses, polymers, and crystals – yields a distribution of local temperatures ($T_x$) which is heterogeneous on the scale of nanometers, unlike the homogenous temperature ($T$) needed for standard thermodynamics [2-5]. Another example is molecular dynamics simulations showing that excess energy fluctuations at low $T$ exceed predictions of standard statistical mechanics by at least an order of magnitude, which can also be characterized by a local $T_x$ that differs from $T$ [6]. In the past, when violations of the 1$^{st}$ law of thermodynamics were found, new terms were added to restore conservation of energy. Examples include energy from the addition of individual molecules as treated by the chemical potential ($\mu$) introduced by Gibbs in 1876, and energy from radioactive decay as treated by its equivalence to mass ($mc^2$) introduced by Einstein in 1905. A lesser-known but similarly crucial contribution to conservation of energy comes from finite-size effects, as treated by the subdivision potential ($\mathcal{E}$) introduced by Hill in 1962 for his thermodynamics of small systems [7,8]. In much of his work, Hill focused on average contributions to $\mathcal{E}$ from surface effects and small-number statistics that can also be treated by including appropriate terms in the Hamiltonian. Here we focus on how this "nanothermodynamics" serves as a guide to treat nanoscale heterogeneity inside macroscopic samples. In general, such finite-size effects involve contributions to energy that do not appear explicitly in the Hamiltonian, instead emerging from the bulk behavior in a way that requires nanothermodynamics for a systematic treatment. Here we review some results that arise by requiring strict adherence to the laws of thermodynamics, even on the scale of nanometers.

The broad generality of thermodynamics can be a challenge to some physics students, who may prefer the concrete models and microscopic details that comprise statistical mechanics. For similar reasons, statistical mechanics is often considered to be a foundation for thermodynamics. Indeed, there are several excellent texts [9-13] and tutorials [14-16] to help explain the power and utility of statistical physics. However, the fundamental physical laws are in the thermodynamics, and strict adherence to these laws is required before statistical mechanics can fully represent the real world. Here we briefly explain how the laws of thermodynamics can be extended to length scales of nanometers, and why applying nanothermodynamics to many models may improve their accuracy and relevance to real systems.

The rest of this overview is organized as follows. In section **II** we briefly recount the history of nanothermodynamics. In section **III**, we **A)** describe some general features of nanothermdynamics, then apply it to **B)** the semi-classical ideal gas, and **C)**-**E)** various forms of Ising-like models for binary degrees of freedom ("spins") in a one-dimensional (1-D) lattice. In section **IV** we conclude with a brief summary.

**II. History of nanothermodynamics**

Classical thermodynamics was originally developed to describe the behavior of large systems, such as steam engines and power plants. In contrast, statistical mechanics is often applied to small systems, such as electrons, atoms, molecules, polymers, and biomolecules. However, most statistical mechanics is still based on the standard thermodynamic limit of ideal (rapid but weak) thermal contact to an effectively infinite heat bath. A key component of nanothermodynamics involves the self-consistent treatment of small systems that are in thermal contact with an ensemble of similar small systems. The primary assumption of nanothermodynamics is that a large ensemble of small systems should behave like a large system. Thus, nanothermodynamics forms a bridge that connects microscopic thermal properties to macroscopic behavior.

The mathematical foundations of nanothermodynamics were developed by Terrell Hill during the years of 1961-64 [17,18]. His results were published in a journal article [19] and two books [7]. Because the books had no length restrictions, he could thoroughly address most modifications to thermodynamics caused by finite-size effects. All that remained was to adapt his foundations to treat the statistical mechanics of specific models. However, Hill abruptly changed his main research interest to focus on molecular biology, and never actively promoted his work. Thus, for more than 35 years (with few exceptions) Hill was the only person to utilize nanothermodynamics. Still today, Hill's crucial extension of thermodynamics to include finite-size effects remains relatively unknown. Partial explanation for this dormancy was recognized by Joel Keizer in 1987 [20]: "It may be, as with much of Terrell's work, that it was simply ahead of its time and that in future years much will be made of it." Now that we are nearing 60 years since Hill first initiated the theory of small system thermodynamics, it is hoped that this basic introduction may help it gain some appreciation.

A common assumption in standard thermodynamics is that if two identical systems are combined, all extensive variables (such as internal energy $E$, entropy $S$, and number of particles $N$) must double, whereas all intensive variables (such as temperature $T$, pressure $P$, and chemical potential $\mu$) must remain unchanged. In fact, the Gibbs-Duhem relation, found by assuming that all extensive variables increase linearly with *N*, is often used as a test of thermodynamic consistency. For example, Gibbs' paradox [21-23] comes from the apparent discrepancy between the predictions of thermodynamics and classical statistical mechanics. Specifically, standard thermodynamics states that when a partition is reversibly removed between two identical systems the total entropy should be exactly twice the entropy of each initial system, while Maxwell-Boltzmann statistics predicts an increase by a term proportional to *N*ln(2). Here we obtain several results where various thermodynamic quantities (including *E*, *S*, and $\mu$) depend nonlinearly on *N*, and we focus on how this dependence may be necessary to maximize the total entropy.

### III. Finite-size effects in the thermal properties of simple systems
#### A. General aspects

Two key variables that form the foundation of thermodynamics are temperature and entropy. Temperature is the familiar quantity that can be defined as "hotness measured on some definite scale" [12], but entropy is a more-subtle concept [24-27]. Although entropy is often associated with randomness, any system with sufficient information about the randomness may also have low entropy. For example, a deck of cards that is fully ordered in its initial sequence of A, 2, 3, 4, … may be said to have zero entropy, but a shuffled deck with a random but known sequence may also have zero entropy; card games rely on the assumption that no one knows the specific sequence of shuffled cards. Thus, a more-general definition of entropy comes from quantifying the amount of *missing* information.

In statistical mechanics, entropy usually involves calculating the number of distinct microscopic configurations that yield the same macroscopic energy. This may be expressed in terms of the multiplicity ($\Omega$) for the number of microstates that yield a specific macrostate, or in terms of their thermal probability ($\rho$). The Gibbs (or Boltzmann-Gibbs) expression for entropy is given by $S/k = -\rho\ln(\rho)$. Here $k$ is Boltzmann's constant, which was first accurately quantified by Max Planck during his development of quantum mechanics [28]. Alternatively, the Boltzmann expression for entropy is given by

$$S/k = \ln(\Omega), \tag{1}$$

Although an equivalent equation is now found on Boltzmann's tomb, he never expressed entropy in this form; it was inscribed by his followers many years after his death [29]. In the microcanonical ensemble, where all microstates are assumed to be equally likely, both expressions for $S$ yield identical values for equilibrium average behavior. (Generalized entropies have been introduced to investigate the possibility that all microstates are not equally likely [30], but here we focus on specific non-additive and nonextensive contributions to entropy that arise without changing the foundations of statistical mechanics.) Gibbs' expression has the advantage that it is also valid in other macroscopic ensembles, while Boltzmann's expression has the advantage that it also applies to non-equilibrium conditions [31,32], including small systems that may fluctuate.

Experimentally, changes in entropy are usually found by measuring the heat capacity ($C$) over a range of temperatures, then integrating $\Delta S = \int dT(C/T)$ [33]. For a gas, these measurements are usually made at fixed total volume ($V$) and constant number of particles ($N$), yielding the canonical ensemble. Theoretically, this entropy is often calculated from the difference between internal energy and Helmholtz free energy ($A$) using: $S/k = (E-A)/(kT)$. However, this expression of entropy assumes that the system is macroscopically homogeneous. In nanothermodynamics, a large system is allowed to subdivide into a heterogeneous distribution of regions with all possible sizes. The resulting heterogeneity and finite-size effects alter the total entropy, yielding the equation:

$$S/k = (E+PV-\mu N-\mathcal{E})/(kT). \tag{2}$$

Note that pressure and temperature are standard (intensive) thermodynamic variables, but that in nanothermodynamics the chemical potential may depend weakly on $N$. The new quantity that is unique to nanothermodynamics is the subdivision potential $\mathcal{E}$, which is added to extensive quantities but always varies nonlinear with the size of the system. One way to understand $\mathcal{E}$ is to compare it to $\mu$. The chemical potential is the change in (free) energy to take a single particle from a bath of particles into the system, whereas $\mathcal{E}$ is the additional change in energy to take a cluster of interacting particles from a bath of clusters into the system, and in general $N$ interacting particles do not have the same energy as $N$ isolated particles, due to surface effects, length-scale terms, finite-size fluctuations, etc. Many finite-size contributions to energy, such as surface states and length-scale quantization, can be included in the net Hamiltonian of the system. Indeed, already in 1872, Gibbs included surface energies proportional to $N^{2/3}$. However, in general it is impossible to fully account for all contributions to energy and entropy without including $\mathcal{E}$, especially from internal fluctuations, and thermal mixtures of distinguishable and indistinguishable particles. Thus, $\mathcal{E}$ is necessary to conserve energy from all sources, especially those on the scale of nanometers.

Thermodynamics establishes equations and inequalities between thermodynamic variables that must be obeyed by nature. In general, the thermodynamic variables come in pairs, whose product yields a contribution to the internal energy. The fundamental equation of thermodynamics (aka the Gibbs equation, which combines the first and second laws) states that the total internal energy of a system can be changed by changing one (or more) of the thermodynamic variables. The original Gibbs equation for a single-component system has three pairs of conjugate variables.

However, nanothermodynamics, adds a fourth pair, the subdivision potential ($\mathcal{E}$) and the number of subdivisions ($\eta$). Thus, the fundamental equation for simple systems becomes [34]:

$$dE = TdS - PdV + \mu dN + \mathcal{E}d\eta \qquad (3)$$

Equation (3) states that there are four ways to change the internal energy (*dE*) of the system: heat can be added (*TdS*), work can be done (*PdV*), particles can be added ($\mu dN$), or the system can be subdivided into an ensemble of smaller regions ($\mathcal{E}d\eta$); or more importantly it can subdivide itself to find its true thermal equilibrium. For ideal gases and other non-interacting systems, $\mathcal{E}$ comes from purely entropic effects. For interacting systems, subdividing into smaller regions adds interfaces that usually have higher energy. However, the energy of these interfaces can be relatively small for large regions, so that reductions in energy and increases in entropy from finite-size effects can favor subdivision. Examples of energy reductions include added surface states, and increased fluctuations towards the ground state. Thus, the subdivision potential provides the only systematic way to maximize total entropy while accommodating contributions to energy that emerge from nanoscale thermal fluctuations, surface states, and waves that are influenced by transient disorder.

The thermodynamic ensemble and resulting free energy that is most relevant to a system depends on how it is coupled to its environment. In general, each pair of conjugate variables that contribute to the total energy of a system includes an "environmental" variable (controlled by the environment, e.g. researcher) and another variable that responds to this control. Different sets of environmental variables form distinct thermodynamic ensembles. Different ensembles may be connected to statistical mechanics using ensemble averages that are equated to thermodynamic quantities. In principle, for simple systems having 3 pairs of conjugate variables there are eight ($2^3$) possible ensembles [35]. In practice, however, only seven of these ensembles are well-defined in standard thermodynamics. Examples include the fully-closed microcanonical ensemble, as well as the partially-open ensembles: canonical, Gibbs', and grand-canonical. The fully-open generalized ensemble, involving three intensive environmental variables (e.g. $\mu$, $P$, $T$) is ill-defined in standard thermodynamics because the size of the system is not fixed, so that nanothermodynamics is required to treat this "nanocanonical" ensemble in a consistent manner. In fact, the nanocanonical ensemble is the only ensemble that does not externally constrain the distribution of sizes inside a bulk sample. Thus, the nanocanonical ensemble is usually required to find the true thermal equilibrium of systems that are comprised of independent internal regions. Because homogeneous systems in the thermodynamic limit all yield equivalent behavior for all ensembles, it is often said that the choice of ensemble can be made purely for mathematical convenience. However, for small systems, and for bulk samples that subdivide into an ensemble of small regions, the choice of ensemble is crucial, so that the correct ensemble must be used for realistic behavior. Indeed, the correct ensemble is essential for fully-accurate descriptions of fluctuations, dynamics, and the distribution of independent regions inside most samples.

Standard statistical mechanics usually starts by calculating a partition function using a simplified model of a physical system. The partition function involves summing (or integrating) over all possible states of the model, weighted by the probability of each state. An example in the canonical ensemble is:

$$Q_{N,V,T} = \sum_E \Omega_{N,V,E} e^{-E/kT} \tag{4}$$

Equation (4) is used to calculate the Helmholtz free energy $A_{N,V,T} = -kT\ln(Q)$ and the average internal energy:

$$\bar{E}_{N,V,T} = \frac{\partial \ln Q}{\partial (-1/kT)} \tag{5}$$

In the canonical ensemble, the chemical potential is fixed by the number of particles, *N*, and is given by:

$$\mu = \frac{\partial A}{\partial N} \tag{6}$$

A second Legendre transform [36] yields the partition function for the grand canonical ensemble:

$$\Xi_{\mu,V,T} = \sum_{E,N} \Omega_{N,V,E} e^{-E/kT} e^{\mu N/kT} \tag{7}$$

which gives the grand potential $\Phi_{\mu,V,T} = -kT \ln(\Xi)$. Because nanothermodynamics includes nonextensive contributions to energy, it allows a third Legendre transform into the nanocanonical ensemble:

$$\Upsilon_{\mu,P,T} = \sum_{N,VE} \Omega_{N,V,E} e^{-E/kT} e^{\mu N/kT} e^{-PV/kT} \tag{8}$$

which yields the subdivision potential $\mathcal{E}_{\mu,P,T} = -kT \ln(\Upsilon)$. Alternatively, the subdivision potential can be found by removing all extensive contributions to the internal energy:

$$\mathcal{E}_{\mu,P,T} = E - TS + PV - \mu N \tag{9}$$

In standard thermodynamics, the Gibbs-Duhem relation requires that $\mathcal{E}_{\mu,P,T}$ must always be zero. Indeed, it is sometimes said that self-consistency in thermodynamics requires $\mathcal{E}_{\mu,P,T} \equiv 0$ so that all contributions to the internal energy are extensive. However, finite-size effects invariably involve nonextensive contributions to entropy and energy. Thus, nanothermodynamics may be thought of as thermodynamics without the artificial restrictions imposed by the Gibbs-Duhem relation.

Most examples presented here are based on the Ising model for binary degrees of freedom ("spins"), with each spin located at a fixed site on a 1-D lattice. The model was originally used by Ernst Ising [37] in an attempt to explain ferromagnetic phase transitions using spins with a magnetic moment *m*. Binary states of the spins come from assuming that they are uniaxial, constrained to point either "up" (*m* in the +z-direction) or "down" (*m* in the –z-direction).

Assuming that the magnetic field ($B$) is in the +z-direction, the energies of the two alignments are $-mB$ and $+mB$, respectively. Ising's model applies equally well to other systems having binary degrees of freedom, such as the interacting lattice gas of occupied or unoccupied sites, or the binary alloy of two types of atoms on a lattice. The standard 1-D model in the thermodynamic limit, solved by Ising in 1925, has no phase transition. Onsager's tour-de-force mathematical treatment of the Ising model in 2-D was the first analytic solution of a microscopic model to show a phase transition. Indeed, the Ising model is the simplest microscopic model having a thermal phase transition, and remains widely studied to investigate how statistical mechanics can be used to yield thermodynamic behavior. However, Onsager's solution assumes an infinite homogenous system in a specific ensemble that may not always apply to real systems. Because the spins are fixed to a rigid lattice, $P$ and $V$ play no role in the energy, replaced by the conjugate variables of the magnetic field and total magnetic moment ($M$). For the simple models presented here, either $B$ will be zero or $E$ will be a simple function of $M$, so that only two environmental variables are needed to define the ensemble. The canonical ensemble partition function can be written as:

$$Q_{N,T} = \sum_E \Omega_{N,E} e^{-E/kT} \qquad (10)$$

Using Eq. (10): $A = -kT\ln(Q)$ gives the Helmholtz free energy, Eq. (5) yields the average internal energy, and Eq. (6) gives the chemical potential. Similarly, the nanocanonical partition function is:

$$\Upsilon_{\mu,T} = \sum_N Q_{N,T} e^{\mu N/kT} \qquad (11)$$

which yields the subdivision potential $\mathcal{E}_{\mu,T} = -kT\ln(\Upsilon)$. Alternatively, the subdivision potential can be found by removing all extensive contributions to internal energy:

$$\mathcal{E}_{\mu,P,T} = E - TS - \mu N \qquad (12)$$

After first describing nanothermodynamics for the ideal gas of noninteracting atoms, subsequent examples will involve this Ising model for binary degrees of freedom that interact with each other, or with an external magnetic field.

### B. Ideal gas

Thermal fluctuations and other finite-size effects are often assumed to negligibly alter the average properties of large systems [38-40]. However, we now show that finite-size effects may be necessary to find the true thermal equilibrium in systems of any size. First focus on a large volume ($V \sim 1$ m$^3$) containing on the order of Avagadro's number of atoms ($N \sim N_A = 6.022 \times 10^{23}$ atoms/mole). Assume monatomic atoms at temperature $T$ with negligible interactions (ideal gas), so that the average internal energy comes only from their kinetic energy, $\bar{E} = 3N(\frac{1}{2}kT)$. Gibbs' paradox [21-23] is often used to argue that the entropy of such thermodynamic systems must be additive and extensive. Nanothermodynamics is based on assuming standard thermodynamics in the limit of large systems, while treating nonextensive contributions to thermal properties of small systems in a self-consistent way. Here we review and reinterpret some results given in chapters 10 and 15 of Hill's Thermodynamics of Small Systems [7] to emphasize that entropy is often sub-additive, so that subdividing a large system into an ensemble of regions increases the net entropy. Thus, nature often favors forming nanometer-sized regions, so that nanothermodynamics is necessary for an accurate description.

Figure 1 is a cartoon sketch showing some ways of subdividing a large system into subsystems. The microcanonical ($N,V,E$) ensemble comes from assuming fully-closed subsystems, surrounded by walls that are impermeable, rigid, and insulating, fully isolating every subsystem from its environment, thereby conserving the number of particles, volume, and energy of each subsystem. The canonical (*N*,*V*,*T*) ensemble comes from assuming subsystems surrounded by walls that are impermeable and solid, but thermally conducting, allowing heat to pass freely in and out, so that energy fluctuates and *T* replaces *E* as an environmental variable. The grand canonical (*μ*,*V*,*T*) ensemble comes from assuming subsystems surrounded by solid diathermal walls that are permeable, allowing particles to exchange freely between regions, so that the number of particles fluctuates and *μ* replaces *N* as an environmental variable. The nanocanonical (*μ*,*P*,*T*) ensemble comes from fully-open *regions*, surrounded by walls that are diathermal, permeable, and flexible, allowing volume to change as the particles and heat pass in and out. Thus, the nanocanonical ensemble removes all external constraints from inside the system, allowing a bulk system to find its equilibrium distribution of internal regions. Table I gives the partition function, fundamental equation, and other thermodynamic expressions for an ideal gas in these four ensembles [41]. Symbols in the table include the thermal de Broglie wavelength $\Lambda = h/\sqrt{2\pi mkT}$, and the absolute activity $\lambda = e^{\mu/kT}$.

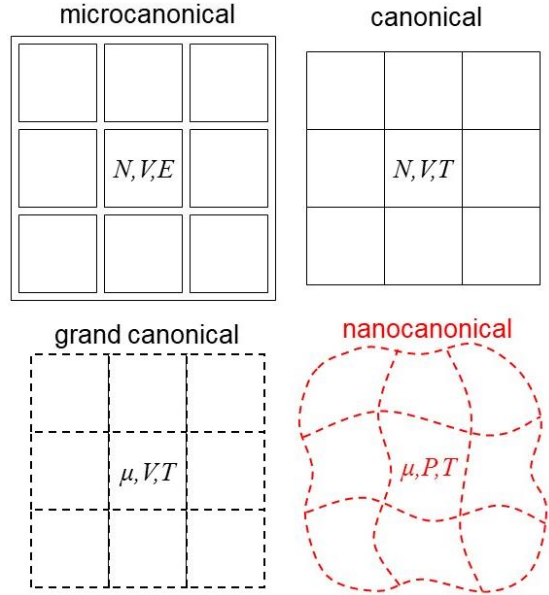

Figure 1. Sketch of various ensembles showing how a large system may subdivide into smaller systems.

Table 1 elucidates some general features of thermodynamics. The microcanonical partition function comes from the multiplicity of microscopic states that have energy *E*. Partition functions in other ensembles come from one or more Legendre transforms to yield other sets of conjugate variables. If the transformation involves a continuous variable, it should be done using an integral over the variable. However, if the variable is discrete (e.g. for the number of atoms *N*), it is

| Table I: monatomic ideal gas |||
|---|---|---|
| Ensemble | Partition function | Fundamental equation and related expressions |
| microcanonical $(N,V,E)$ | $\Omega_1 = V \frac{\pi}{4}\left(\frac{8m}{h^2}\right)^{3/2}\sqrt{E}$<br>$\Omega_N \approx \frac{1}{N!}\left[V\left(\frac{4\pi mE}{3Nh^2}e\right)^{3/2}\right]^N$ | $S/k = \ln(\Omega_N) \approx \frac{3}{2}N + N\ln(V) + \frac{3}{2}N\ln\left[\frac{4\pi mE}{3Nh^2}\right] - \ln(N!)$<br>$\approx \frac{5}{2}N - N\ln\left(\frac{N}{V}\right) + \frac{3}{2}N\ln\left[\frac{4\pi mE}{3Nh^2}\right] - \ln(\sqrt{2\pi N})$ |
| canonical $(N,V,T)$ | $Q_1 = \int_0^\infty \Omega_1 e^{-\frac{E}{kT}}dE = \frac{V}{\Lambda^3}$<br>$\Lambda = h/\sqrt{2\pi mkT}$ $Q_N = \frac{Q_1^N}{N!}$ | $A/kT = -\ln(Q_N) = -N\ln(Q_1) + \ln(N!) \approx N\ln(N\Lambda^3/V) - N + \ln(\sqrt{2\pi N})$<br>$\tilde{\mu}/kT \equiv -\ln(Q_{N+1}/Q_N) = \ln(\Lambda^3/V) + \ln(N+1)$ $\bar{E} = \frac{\partial \ln Q_N}{\partial(-1/kT)} = \frac{3}{2}NkT$<br>$S/k = (\bar{E}-A)/kT \approx 5N/2 - N\ln(N\Lambda^3/V) - \ln(\sqrt{2\pi N})$ |
| grand canonical $(\mu,V,T)$ | $\Xi = \sum_{N=0}^\infty \frac{1}{N!}\left[\frac{V}{\Lambda^3}\right]^N e^{\frac{\mu N}{kT}}$ | $\Phi/kT = -\ln(\Xi) = -V\lambda/\Lambda^3$ $\lambda = e^{\mu/kT}$<br>$\mu/kT = \ln(\bar{N}\Lambda^3/V)$ $\bar{N} = -\partial\Phi/\partial\mu = V\lambda/\Lambda^3$<br>$S/k = (\bar{E} - \Phi - \mu\bar{N})/kT = 5\bar{N}/2 - \bar{N}\ln(\bar{N}\Lambda^3/V)$ |
| nanocanonical $(\mu,P,T)$ | $\Upsilon = \int_0^\infty e^{\frac{V}{\Lambda^3}\lambda}e^{-\frac{PV}{kT}}d\left[\frac{pV}{kT}\right]$ | $\mathcal{E}/kT = -\ln(\Upsilon) = \ln[1-kT\lambda/P\Lambda^3] = -\ln[\bar{N}+1]$<br>$\bar{N} = \lambda\partial\ln(\Upsilon)/\partial\lambda = (kT\lambda/P\Lambda^3)/[1-kT\lambda/P\Lambda^3]$<br>$S/k = (\bar{E} + P\bar{V} - \mu\bar{N} - \mathcal{E})/kT = 5\bar{N}/2 - \bar{N}\ln(\bar{N}\Lambda^3/\bar{V}) + \ln(\bar{N}+1)$ |

especially important in nanothermodynamics to use a discrete summation, thereby maintaining accuracy down to individual atoms, which also often simplifies the math and removes Stirling's approximation for the factorials. Similarly, note that the chemical potential in the canonical ensemble is calculated using a difference equation, not a derivative, so that again Stirling's approximation can be avoided. Another general feature to be emphasized is that the variables shown in the "Ensemble" column are fixed by the environment (e.g. types of walls surrounding a subsystem); hence they do not fluctuate. In contrast, each conjugate variable fluctuates due to contact with the environment, so that these conjugate variables are shown as averages. Thus, as should be expected for small systems, it is essential to use the correct ensemble for determining which variables fluctuate, and by how much.

Now focus on the entropy. Recall that the Sackur-Tetrode formula for the entropy of an ideal gas, introduced in 1912 and now standard in most textbooks, is $S_0/k = 5N/2 - N\ln[N\Lambda^3/V]$. Note that to make this entropy extensive, the partition function is divided by $N!$, which assumes that all the atoms in the system are indistinguishable, usually attributed to quantum symmetry. However, the need to use quantum mechanics for macroscopic systems of semi-classical particles remains a topic of debate [21-23]. Table I shows that nanothermodynamics includes nonextensive contributions to entropy by subtracting the subdivision potential $S/k = S_0/k - \mathcal{E}/kT$. For example, in the canonical ensemble $\mathcal{E}/kT \approx \ln\sqrt{2\pi N}$, which comes from using Stirling's approximation for $N!$. Because the Legendre transform from $N$ to $\mu$ is done by a discrete sum over all $N$, Stirling's approximation is eliminated from the grand-canonical and nanocanonical ensembles. Instead, a novel nonextensive contribution to entropy arises in the nanocanonical ensemble from $\mathcal{E}/kT = -\ln(\bar{N}+1)$. Because this negative subdivision potential is subtracted from $S/k$, the entropy per particle increases when the system subdivides into smaller regions. However, this entropy increase appears only in the nanocanonical ensemble, where the sizes of the regions are unconstrained, a feature that is unique to nanothermodynamics. Figure 2 is a sketch of how net entropy changes if a single system subdivides into subsystems: decreasing if subsystems are constrained to have fixed $V$ and $N$ (canonical ensemble), but increasing if subsystems have variable $V$ and $N$ (nanocanonical ensemble). Such subadditivity found in nanocanonical systems is also a requirement of quantum-mechanical entropy [42].

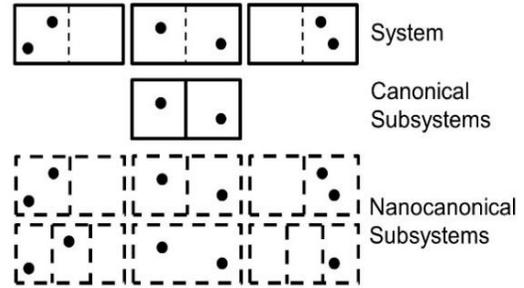

Figure 2. Crude characterization of a system (top row) and its multiplicities for two types of subdivision. The particles (dots) can be on either side of the system, but must be in separate volumes for fixed $N$ in canonical subsystems (second row). Nanocanonical subsystems have variable $N$, and variable $V$, increasing their entropy.

The subdivision potentials from Table I can be used to obtain the nonextensive corrections to entropy of specific ideal gases in various ensembles. As an example, consider one mole ($N$=6.022x10$^{23}$ atoms) of argon gas (mass $m$=6.636x10$^{-26}$ kg) at a temperature of 0 °C ($T$=273.15 K), yielding the thermal de Broglie wavelength $\Lambda = h/\sqrt{2\pi mkT} \approx 16.7$ pm. At atmospheric pressure (101.325 kPa) the number density is one amagat ($N/V$ = 2.687x10$^{25}$ atoms/m³), yielding an average distance between atoms of $(V/N)^{1/3} = (\bar{V}/\bar{N})^{1/3} \approx 3.34$ nm, and mean-free path of $\ell = (V/N)/(\sqrt{2}\pi d^2) \approx 59.3$ nm (using $d=0.376$ nm as the kinetic diameter of argon). Under these conditions the Sackur-Tetrode formula predicts a dimensionless entropy per atom of $S_0/Nk = 5/2 - \ln[\Lambda^3 \bar{N}/\bar{V}] \approx 18.39$ (equal to 152.9 J/mole-K). In the canonical ensemble the subdivision potential is positive, $\mathcal{E}/kT \approx \ln(\sqrt{2\pi N})$, so that when subtracted from the Sackur-Tetrode formula the entropy is reduced. Although the magnitude of the entropy reduction per atom is microscopic, $\mathcal{E}/NkT = 4.70 \times 10^{-23}$, any such reduction is enough to support the standard thermodynamic hypothesis of a single homogeneous system. However, the hypothesis fails if subsystems are not explicitly constrained to have a fixed size. Indeed, *regions* in the nanocanonical ensemble have a subadditive entropy that increases upon subdivision. Specifically, $\mathcal{E}/kT = -\ln(\bar{N} + 1)$ is negative when $\bar{N} > 0$, confirming that any system of ideal gas atoms favors subdividing into an ensemble of regions whenever the size of each region is not externally constrained. Thermal equilibrium in the nanocanonical ensemble is usually found by setting $\mathcal{E}=0$, yielding $\bar{N} \to 0$ and an increase in entropy per atom of: $-\mathcal{E}/\bar{N}kT = \lim_{\bar{N}\to 0} [\ln(\bar{N}+1)]/\bar{N} = 1$, about 5.4 % of the Sackur-Tetrode component. However, the Sackur-Tetrode formula has been found to agree with measured absolute entropies of four monatomic gases, with discrepancies (0.07-1.4 %) that are always within two standard deviations of the measured values [33]. Thus, the experiments indicate that $\bar{N} \gg 1$ in real gases, presumably due to quantum symmetry on nanometer length scales. For example, assume that quantum symmetry (indistinguishability) occurs for atoms over an average distance of the mean-free path, so that $\bar{N} = \ell^3 (N/V) \approx 5600$ atoms. Now the subdivision potential per atom yields $-\mathcal{E}/\bar{N}kT = [\ln(\bar{N}+1)]/\bar{N} \approx 0.0015$, well within the experimental uncertainty. In any case, maximum total entropy should always be favored, no matter how small the gain. Thus, quantum symmetry may apply to ideal gas atoms at normal temperatures and pressures across nanometer-sized regions, but not across macroscopic volumes.

Having $\bar{N} \ll N$ for a semi-classical ideal gas implies that many atoms can be distinguished by their local region within the large system. Thus, as expected, a large system of indistinguishable atoms can increase its entropy by making most of the atoms distinguishable by their location. Furthermore, because the nanocanonical ensemble allows fluctuations around $\bar{V}$, regions may adapt their size to encompass atoms that are close enough to collide, or at least to have wavefunctions that may overlap, which is a common criterion for signaling the onset of quantum behavior.

Figure 3 depicts the entropy of mixing for an ideal gas with identical particles (upper-left boxes), or different particles (lower-left boxes). In the canonical ensemble (upper equations), extensive entropy is approximated (terms containing $\ln\sqrt{2\pi N}$ are neglected) by assuming macroscopic systems of indistinguishable particles, as discussed in many textbooks. Here we focus on the nanocanonical ideal gas (lower equations). Recall that the nonextensive term maximizes the entropy when $\bar{N} \ll N$, so that the volume of each region is microscopic and variable, i.e. *not* the size of the container. First consider doubling the size of a single system containing a single type of atom (same-color boxes in upper-left). The final state (right side) has $V'=2V_1$ and $N_1'=2N_1$ (with $N_2=N_2'=0$). Whereas, the average number of atoms per region does not change $\bar{N}_1' = \bar{N}_1$, nor does their density $\bar{N}_1'/\bar{V}_1' =$ $\bar{N}_1/\bar{V}_1$. Thus, the main contribution to the total entropy of the combined system is simply twice the entropy of each initial system: $S_1'/k \approx \frac{5}{2}N_1' - N_1' \ln[\bar{N}_1' \Lambda^3/\bar{V}_1'] \approx 2S_1/k$, in agreement with the Sackur-Tetrode formula. However, because $\bar{N}_1' = \bar{N}_1$ in the nanocanonical ensemble (subscript *nc*), the final entropy of the combined system is lower than the initial entropy of two identical subsystems from the difference: $\Delta S_{nc}/k = \ln(\bar{N}_1' + 1) - 2\ln(\bar{N}_1 + 1) = -\ln(\bar{N}_1 + 1)$. It is this subadditive reduction in entropy as systems combine that favors subdividing bulk systems into nanoscale regions, but only in the nanoncanonical ensemble where there are no external constraints on the sizes of the regions. In contrast, combining identical subsystems in the canonical ensemble increases the total entropy. Specifically, in the canonical ensemble (subscript *c*) the positive subdivision potential causes a net increase in entropy for the combined system of $\Delta S_c/k = -\ln(\sqrt{2\pi 2N_1}) + 2\ln(\sqrt{2\pi N_1}) = \ln(\sqrt{\pi N_1})$, a large positive quantity if $N_1$ is on the order of Avagadro's number, but still small compared to the extensive entropy terms that depend linearly on $N_1$. Such non-extensive contributions to entropy are usually neglected in standard thermodynamics, where all ensembles are assumed to yield equivalent results. In nanothermodynamics, however, the subadditive contribution from the subdivision potential in the nanocanonical ensemble favors subdividing into an ensemble of small regions, regardless of the size of the initial system, fundamentally different from the single macroscopic system favored by the canonical ensemble.

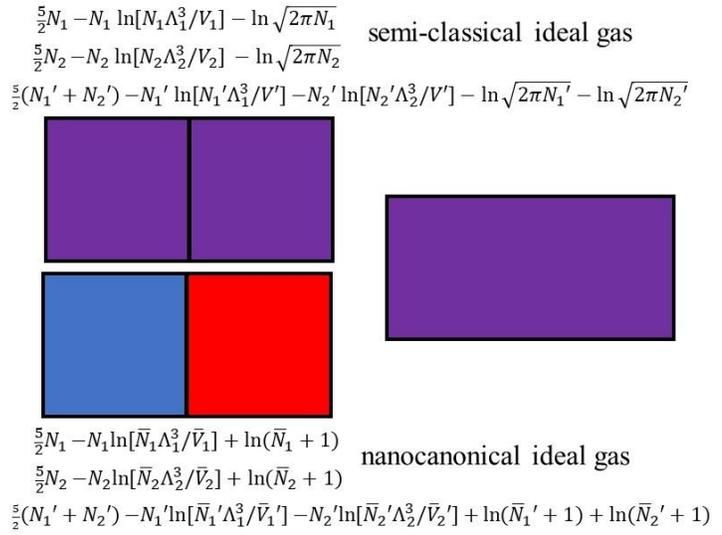

Figure 3. Entropy of mixing from combining two similar systems of an ideal gases (upper-left), or two dissimilar ideal gases (lower-left). The dominant (extensive) contributions to the entropies of mixing are equivalent for ideal gases in the canonical ensemble (upper equations) and nanocanonical ensemble (lower equations), but the nonextensive terms have opposite signs.

Next consider the case of combining two identical volumes ($V_1=V_2=V_1'/2$) having different types of atoms: $N_1'=N_1$ and $N_2'=N_2$ with $N_1=N_2$. The initial state is represented by the lower-left diagram in Fig. 3 showing two systems with different colors. Now, because the volume is doubled without adding new atoms of either type, the final densities are half their initial values, $\bar{N}_1'/\bar{V}_1' = N_1'/V_1' = N_1/(2V_1) = \frac{1}{2}\bar{N}_1/\bar{V}_1$ and $\bar{N}_2'/\bar{V}_1' = \frac{1}{2}\bar{N}_2/\bar{V}_2$. Thus, the main contribution to the final entropy is $S'/k \approx \frac{5}{2}(N_1 + N_2) - N_1 \ln[\frac{1}{2}\bar{N}_1 \Lambda^3/\bar{V}_1] - N_2 \ln[\frac{1}{2}\bar{N}_2 \Lambda^3/\bar{V}_2]$; greater than the initial entropy by an amount $\Delta S = (N_1 + N_2)\ln(2)$. Indeed, this dominant term in the entropy of mixing in the nanocanonicial ideal gas matches the behavior of the Sackur-Tetrode formula. Thus, nanothermodynamics allows an ideal gas to maximize its entropy, and meet the requirements of entropy of mixing, without resorting to macroscopic quantum behavior for ideal gas particles that may be meters apart, and hence distinguishable by their location. However, because the nanocanonical ensemble allows the number of atoms in a particular region to fluctuate, the number of indistinguishable atoms in a specific local region may often be $N>1$ when atoms are close enough to collide, or when their de Broglie wave functions might entangle. Thus, a novel solution to Gibbs' paradox comes from including nonlinear size-dependent terms in the entropy of an ideal gas, without requiring quantum symmetry conditions for macroscopic systems. This fundamental result stresses the importance of finite-size contributions to thermal properties, which can be fully understood in a general manner only through nanothermodynamics.

### C. Finite chain of Ising spins

Simple models of magnetic spins provide a basic scenario for studying finite-size effects in thermal systems. The fundamental equation of thermodynamics for magnetic systems, including the subdivision potential from nanothermodynamics, is given in Fig. 4. Also shown is a set of cartoon sketches of energy-level diagrams indicating how the various contributions change the internal energy. Each sketch shows three energy levels, with dots depicting the relative occupation of each level. The occupation of these levels for an initial internal energy is shown by the left-most energy-level diagram. The next three energy-level diagrams, from left-to-right respectively, show that: adding heat (*TdS*) alters the relative occupation of all levels, doing magnetic work (*–MdB*) changes the spacing of the levels, while adding spins (*μdN*) increases the occupation of each level. The right-most energy-level diagram represents the change in energy from the subdivision potential (*Ɛdη*). When the number of subdivisions is increased (*dη*>0) the average size of small regions is reduced, allowing energy levels to be broadened by finite-size effects in small regions due to surface states, interfaces, thermal fluctuations, etc. The subdivision potential, which is unique to nanothermodynamics, allows systematic treatment of these finite-size effects, thereby ensuring that energy is strictly conserved, even on the scale of nanometers.

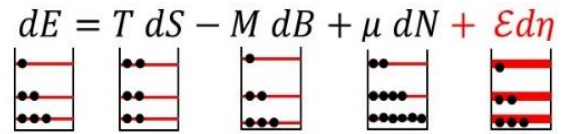

Figure 4. Fundamental equation for conservation of energy in magnetic systems, including finite-size effects, with a sketch of how a three energy-level system can be changed by various contributions.

The Ising model for uniaxial spins (binary degrees of freedom) demonstrates the power and utility of nanothermodynamics for finding the thermal equilibrium of finite-sized systems. Exact results can be obtained analytically in one dimension (1-D) in zero magnetic field (*B*=0). Assume *N* Ising spins, each having magnetic moment *m* that can be aligned in the +*z* or –*z* direction, with interactions only between nearest-neighbor spins. Let the spins favor ferromagnetic alignment, so that the energy of interaction (exchange energy) is –*J* if the two neighboring spins are in the same direction, and +*J* if they are anti-aligned. The usual solution to the 1-D Ising model includes contributions to energy from the external magnetic field and the exchange interaction, yielding a partition function [9,11]:

$$Z(N,B,T) \approx \{e^{J/kT}\cosh(mB/kT) + [e^{-2J/kT} + e^{2J/kT}\sinh^2(mB)\,]^{1/2}\}^N \qquad (11)$$

If *B*=0, the resulting free energy becomes:

$$F(N,0,T) = -kT\ln Z(N,0,T) \approx -NkT\ln[2\cosh(J/kT)] \qquad (12)$$

The approximations come from finite-size effects that have been neglected in Eqs. (11) and (12) by assuming large systems, with periodic boundary conditions (or equivalently spins in a ring).

However, most real spin systems do not form rings, and with finite-size effects that are suppressed by periodic boundary conditions. Thus, for most realistic spin chains Eqs. (11) and (12) are strictly valid only in the limit of large systems, $N \to \infty$. We now consider finite-size effects explicitly.

Let there be $N+1$ spins in a linear chain, yielding a total of $N$ interactions ("bonds") between nearest-neighbor spins [43]. It is convenient to write the energy in terms of the binary states of each bond, $b_i = \pm 1$. Using $+J$ for the energy of anti-aligned neighboring spins, and $-J$ between aligned neighbors. The Hamiltonian is:

$$E = -J \sum_{i=1}^{N} b_i, \text{ with } b_i = \{-1, +1\}. \tag{13}$$

Assuming $x$ high-energy bonds, so that there are $(N-x)$ low-energy bonds, the internal energy is $E = -J(N-2x)$. The multiplicity of ways that this energy can occur is given by the binomial coefficient. The thermal properties in various ensembles are given in Table II. Note that due to end effects, the Helmholtz free energy from Table II for $N$ spins ($N-1$ bonds) is $A = -(N-1)kT \ln[2\cosh(J/kT)]$, which equals Eq. (12) only when $N \to \infty$. Thus, if there is a macroscopic number of spins in an unbroken chain, the differences are negligible. However, if the length of the chain can change by adding or removing spins at either end, thermal equilibrium requires the nanocanonical ensemble. As with the ideal gas, this nanocanonical ensemble is the only ensemble that does not externally constrain the sizes of the regions, so that the system itself can find its equilibrium average size and distribution of sizes. Theoretically the average size is determined by setting the subdivision potential to zero. For the Ising model, this yields an average number of spins in each chain of: $\bar{N} + 1 = \cosh(J/kT) + 1$. Thus, at high temperatures the average chain contains two spins connected by one bond, whereas when $T\rightarrow 0$ the average chain length diverges.

Table II: $N+1$ Ising spins ($N$ bonds) in zero field with $x$ high-energy bonds ($+J$) and ($N-x$) low-energy bonds ($-J$)

| Ensemble | Partition function | Fundamental equation and related expressions |
|---|---|---|
| microcanonical ($N+1, x$) | $\Omega = \dfrac{2N!}{x!(N-x)!}$ | $S/k = \ln(\Omega) = \ln[N!] + \ln(2) - \ln[x!] - \ln[(N-x)!]$ <br> $\approx N\ln(N) - x\ln(x) - (N-x)\ln(N-x) - \ln[\sqrt{\pi x(1-x/N)/2}\,]$ |
| canonical ($N+1, T$) | $Q = \sum_{x=0}^{N} \Omega\, e^{(N-2x)\frac{J}{kT}}$ | $A/kT = -\ln(Q) = -\ln[2(e^{J/kT} + e^{-J/kT})^N]$     $\mu/kT = -\ln[2\cosh(J/kT)]$ <br> $\quad = -N\ln[2\cosh(J/kT)] - \ln(2)$     $\bar{E}/J = (2\bar{x} - N) = -N\tanh(J/kT)$ <br> $S/k = (\bar{E} - A)/kT = -N(J/kT)\tanh(J/kT) + N\ln[2\cosh(J/kT)] + \ln(2)$ |
| nanocanonical ($\mu, T$) | $\Upsilon = \sum_{N=0}^{\infty} Q\, e^{\frac{\mu(N+1)}{kT}}$ | $\mathcal{E}/kT = -\ln(\Upsilon) = \ln[\tfrac{1}{2}e^{-\mu/kT} - \cosh(J/kT)]$     $\mu/kT = -\ln[2(\bar{N}+1)]$ <br> $\quad = 0$     $\bar{N} + 1 = \tfrac{1}{2}e^{-\mu/kT} = \cosh(J/kT) + 1$ <br> $S/k = (\bar{E} - \mu\bar{N} - \mathcal{E})/kT = -\bar{N}(J/kT)\tanh(J/kT) + \bar{N}\ln[2 + 2\cosh(J/kT)]$ |

As expected, Table II shows that the entropy of Ising spins increases with decreasing constraints, so that again (as in Table I) the nanocanonical ensemble has the highest entropy. Specifically, the entropy per bond in the nanocanonical ensemble is larger than that in the canonical ensemble by the difference $\Delta S/\bar{N}k = \ln[(\bar{N}+1)/\bar{N}] - \ln(2)/\bar{N}$. At high $T$ where $\bar{N}\rightarrow 1$, $\Delta S/\bar{N}k\rightarrow 0$. At low $T$ where $\bar{N}>>1$, $\Delta S/\bar{N}k \approx 1/\bar{N} - \ln(2)/\bar{N}\rightarrow 0$. Numerical solution yields a maximum entropy difference of nearly 6% ($\Delta S/\bar{N}k$=0.0596601…) at $kT/J$ = 0.687297… where $\bar{N}$=2.25889… Hence, Ising spins in the nanocanonical ensemble always have higher entropy than if they were constrained to be in the canonical ensemble, but the excess is small at both low, and high $T$. Nevertheless, the second law of thermodynamics asserts that nature will do whatever is possible to maximize the total entropy, no matter how small the increase. Thus, if a mechanism exists to change the length of the system, an infinite chain will shrink until there is on average $\bar{N} + 1 = \cosh(J/kT) + 1$ spins in each region. In fact, because it can be difficult to fix the size of many small systems, their size should vary without external constraints, limiting the usefulness of the canonical ensemble for describing finite-size effects inside most real systems.

A key feature of the nanocanonical ensemble is that thermal equilibrium is found by setting the subdivision potential to zero. Indeed, setting $\mathcal{E}$=0 allows the system to find its own equilibrium distribution of regions, without external constraint, similar to how $\mu$=0 in standard statistical mechanics yields the equilibrium distribution of phonons and photons, without external constraint. However, $\mathcal{E} = -kT\ln(\Upsilon) = 0$ requires $\Upsilon$=1, so that all relevant factors must be carefully included in the partition function. For example, suppose that the factor of 2 in the numerator of $\Omega$ (Table II) is ignored, from neglecting the degeneracy of a sequence of spins and its inversion. Although averages in the canonical ensemble that are normalized by the partition function do not change, the average number of bonds in the nanocanonical ensemble becomes $\bar{N} = 2\cosh(J/kT)$, twice the value of $\bar{N} = \cosh(J/kT)$ from Table II.

### D. The subdivided Ising model: Ising-like spins with a distribution of neutral bonds

Results similar to those in section **III C)** for the finite-sized Ising model in the nanocanonical ensemble can be obtained in the canonical ensemble by modifying the Ising model to include "neutral bonds," from nearest-neighbor spins that do not interact [44]. Again, start with the standard Ising model having $N+1$ spins ($N$ bonds), but now let there be $\eta$ neutral bonds, in addition to $x$ high-energy bonds between anti-aligned spins, leaving $N$–$\eta$–$x$ low-energy bonds between aligned spins. Figure 5 shows a specific configuration of 11 spins ($N=10$ bonds) with $x=2$ high-energy bonds (**X**), $\eta=3$ neutral bonds (**O**), and $N$–$\eta$–$x=5$ low-energy bonds (●). Physically, neutral bonds may come from neighboring spins with wavefunctions having negligible overlap, or from neighboring spins that fluctuate at different frequencies so that their time-averaged interaction is zero. It is again convenient to write the Hamiltonian in terms of the bonds, which may now have three different states:

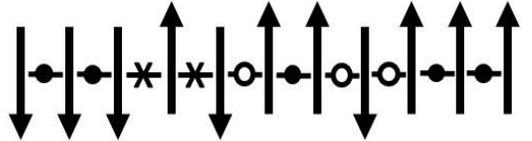

Figure 5. Sketch of 11 Ising-like spins in a chain connected by $N=10$ bonds. Four bonds are low energy (●), two bonds are high-energy (**X**), and three bonds are neutral (**O**).

$$E = -J \sum_{i=1}^{N} b_i, \text{ with } b_i = \{-1, 0, +1\}. \tag{14}$$

The internal energy for the system with fixed $x$ and $\eta$ is: $E = -J(N$–$\eta$–$2x)$. The canonical ensemble involves two sums. The first sum is over $x$ for fixed $\eta$, with a multiplicity given by the trinomial coefficient for the number of ways that the high- and low-energy bonds can be arranged among $N$–$\eta$ interacting bonds. An extra factor of $2^\eta$ arises because each neutral bond has two possible states for its neighboring spin. This first sum yields a type of canonical ensemble for the system with fixed $\eta$. A second sum is over all values of $\eta$. The multiplicity is given by the binomial for the number of ways that the neutral bonds can be distributed, which arises from the trinomial after the first sum. The behavior of this model is summarized in Table III.

| Ensemble | Partition function | Fundamental equation and related expressions |
|---|---|---|
| microcanonical ($N+1,\eta,x$) | $\Omega = \dfrac{2N!\,2^{\eta}}{x!\,\eta!\,(N-\eta-x)!}$ | $S/k = \ln(\Omega) = \ln[N!] + \ln(2^{\eta+1}) - \ln[x!] - \ln[\eta!] - \ln[(N-\eta-x)!]$ <br> $\approx N\ln(N) - x\ln(x) - \eta\ln(\eta/2) - (N-\eta-x)\ln(N-\eta-x)$ |
| quasi-canonical ($N+1,\eta,T$) | $q = \sum_{x=0}^{N-\eta} \Omega\, e^{(N-\eta-2x)\frac{J}{kT}}$ | $q = 2^{N+1}[\cosh(J/kT)]^{N-\eta} N!/[\eta!\,(N-\eta)!]$ <br> $N - \eta - 2\bar{x} = \partial \ln q / \partial (J/kT) = (N-\eta)\tanh(J/kT)$ <br> $\bar{E}/J = -(N-\eta-2\bar{x}) = -(N-\eta)\tanh(J/kT)$ |
| canonical ($N+1,T$) | $Q = \sum_{\eta=0}^{N} \dfrac{N!}{\eta!\,(N-\eta)!}\, q$ | $A/kT = -\ln(Q) = -(N+1)\ln(2) - N\ln[1+\cosh(J/kT)]$ <br> $N - \bar{\eta} = \cosh(J/kT)\,\partial \ln Q/\partial \cosh(J/kT) = N\cosh(J/kT)]/[1+\cosh(J/kT)]$ <br> $\bar{E}/J = -(N-\bar{\eta})\tanh(J/kT) = -N\sinh(J/kT)/[1+\cosh(J/kT)]$ <br> $S/k = -N(J/kT)\sinh(J/kT)/[1+\cosh(J/kT)] + N\ln[1+\cosh(J/kT)]$ |

Table III: $N+1$ Ising-like spins with $\eta$ neutral bonds (0), $x$ high-energy bonds ($+J$), and ($N$-$x$-$\eta$) low-energy bonds ($-J$)

We now compare the results in Table III for the subdivided Ising model with those from Table II for the finite-sized Ising model. In the canonical ensemble, the average energy of the subdivided system is higher (not as negative) as that of the finite system, as expected when neutral bonds replace an equilibrium mixture of mostly low-energy bonds. However, the average energy per interacting bond ($\bar{E}$ from Table III, divided by $N - \bar{\eta}$) is $-J\tanh(J/kT)$, the same as $\bar{E}/N$ from Table II. Another similarity comes from the average number of spins in each region ($\bar{n}$). Using the average number of subdivisions:

$$\bar{\eta} = N/[1+\cosh(J/kT)]. \tag{15}$$

yields:

$$\bar{n} = \frac{N+1}{\bar{\eta}+1} = \frac{1+\cosh(J/kT)}{1+[\cosh(J/kT)]/(N+1)}. \tag{16}$$

Hence, in the limit of $N\to\infty$, Eq. (16) yields $\bar{n} \approx 1 + \cosh(J/kT)$ for the subdivided Ising model in the canonical ensemble, approaching $\bar{N} + 1 = 1 + \cosh(J/kT)$ from Table II for the finite-size Ising model in the nanocanonical ensemble. Indeed, the theory of small-system thermodynamics in section **III C)** involving a large ensemble of small systems is equivalent to the concept of nanothermodynamics (here, section **III D)** involving repeatedly subdividing a system, provided the initial system is sufficiently large. Furthermore, two models with distinct Hamiltonians – Eq. (13) in **III C)** for a system of 2-state bonds and Eq. (14) in **III D)** for a system of 3-state bonds – may yield equivalent results. However, equivalence requires that the correct ensemble is used for each system, nanocanonical for Eq. (13) and canonical for Eq. (14). Thus, the choice of ensemble is crucial for obtaining accurate behavior, even for systems that are in thermal equilibrium in the thermodynamic limit.

A similar subdivided Ising model has been adapted to three dimensions by using mean-field energies instead of instantaneous interactions. Mean-field energies may come from microcanonical averaging of weakly coupled regions, consistent with the thermodynamic heterogeneity that is measured in most materials [2-5]. The resulting "mean-field cluster" model yields excellent agreement with the measured susceptibility of many ferromagnetic crystals [45,8]. The agreement includes measured corrections to scaling throughout the paramagnetic phase, extending deep into the non-classical critical scaling regime close to the Curie transition. Thus, by treating classical mean-field theory [46] in its equilibrium nanocanonical ensemble there is no need to use renormalization group theory [47] for the non-classical critical scaling that is measured in most ferromagnetic materials.

### E. Finite chain of non-interacting Ising-like spins in a magnetic field

We now consider Ising-like spins that interact with an external field $B$, but not with each other. Again, consider $N$ spins, each having magnetic moment $m$. Let $B$ be in the $+z$ direction, and assume $u$ spins are "up" (in the $+z$ direction), with $N$–$u$ spins "down" (in the $–z$ direction). The model is equivalent to noninteracting particles adsorbed on a lattice (ideal lattice gas), with an attractive energy $–2mB$ for each particle on the lattice, as treated in [7]. (The dimensionality of the lattice is irrelevant because there are no interactions between particles.) The internal energy of the system is $E = (N–2u)\,mB$. The multiplicity of ways that the system can have a given alignment is given by the binomial coefficient. The results for various ensembles are summarized in Table IV.

| Table IV: $N$ non-interacting Ising-like spins, each having magnetic moment $m$, with $u$ spins in direction of external field $B$ | | |
|---|---|---|
| Ensemble | Partition function | Fundamental equation and related expressions |
| microcanonical ($N,u$) | $\Omega = \dfrac{N!}{u!\,(N-u)!}$ | $S/k = \ln(\Omega) = \ln[N!] - \ln[u!] - \ln[(N-u)!]$ <br> $\approx N\ln(N) - u\ln(u) - (N-u)\ln(N-u) - \ln[\sqrt{2\pi x(1-u/N)}]$ |
| canonical ($N,T$) | $Q = \sum_{u=0}^{N} \Omega\, e^{(2u-N)\frac{mB}{kT}}$ | $A/kT = -\ln(Q) = -\ln[(e^{mB/kT} + e^{-mB/kT})^N]$    $\bar{u}/kT = -\ln[2\cosh(mB/kT)]$ <br> $= -N\ln[2\cosh(mB/kT)]$    $\bar{U}/mB = (N-2\bar{u}) = -N\tanh(mB/kT)$ <br> $S/k = (\bar{U}-A)/kT = (N+1)\ln(N+1) - (\bar{u}+1)\ln(\bar{u}+1) - (N-\bar{u})\ln(N-\bar{u})$ |
| nanocanonical ($\mu,T$) | $\Upsilon = \sum_{N=0}^{\infty} Q\, e^{\frac{\mu N}{kT}}$ | $\mathcal{E}/kT = -\ln(\Upsilon)$    $\bar{N} = e^{\mu/kT}2\cosh(mB/kT)/[1 - e^{\mu/kT}2\cosh(mB/kT)]$ <br> $= -\ln[\bar{N}+1]$    $\mu/kT = -\ln[2\cosh(mB/kT)] - \ln[(\bar{N}+1)/\bar{N}]$ <br> $S/k = (\bar{U} - \mu\bar{N} - \mathcal{E})/kT = (\bar{N}+1)\ln(\bar{N}+1) - \bar{u}\ln(\bar{u}) - (\bar{N}-\bar{u})\ln(\bar{N}-\bar{u})$ |

Note various similarities in the behavior of this system of noninteracting spins, summarized in Table IV, and the ideal gas of noninteracting atoms, summarized in Table I. In the microcanonical ensemble, both systems have a nonextensive contribution to entropy (due to Stirling's approximation for the factorials) that yields a positive subdivision potential, specifically for the spin system $\mathcal{E}/kT \approx \ln[\sqrt{2\pi u(1-u/N)}]$. As with the ideal gas, this positive subdivision potential inhibits the subdivision of a large system into smaller subsystems. Also like the ideal gas, the factorials are removed from other ensembles by summing over all states. Indeed, in the canonical ensemble here, and in the grand-canonical ensemble for the ideal gas, the free energies are exact (no mathematical approximations) and extensive (no subdivision potential), a consequence of no interface terms and no end-effects in these noninteracting systems. In the nanocanonical ensemble both systems have the same subdivision potential, $\mathcal{E}/kT = -\ln[\bar{N}+1]$, which is always negative for $\bar{N} > 0$. In fact, due to contributions from the nonextensive subdivision potential, the entropy per particle of both systems is maximized by $\bar{N} \to 0$ at all $T$. In other words, the total entropy of a system of non-interacting spins is increased by subdividing into an ensemble of separate regions, but only if there is no restriction on the size of each region. This is a consequence of the entropy per particle of a single large system being lower than the entropy of an ensemble of small systems. In general, if nature can find a way to subdivide large systems into separate regions, a heterogeneous ensemble of independent regions is the true thermal equilibrium.

## IV. Conclusions

The theory of small-system thermodynamics is needed to ensure conservation of energy and maximum entropy in the thermal and dynamic properties of small systems, especially on the scale of nanometers. Here we have emphasized that this "nanothermodynamics" is also crucial for obtaining the true thermal equilibrium of large systems that may subdivide into an ensemble of independently fluctuating subsystems that we call "regions." In general, entropy can only be maximized if these regions have no external constraints; a unique feature of the "nanocanonical" ensemble that is well-defined only in nanothermodynamics. Examples are given from several simple models that are suitable for introductory courses in statistical physics.

One result (section **III B**) is that a semi-classical ideal gas of indistinguishable atoms favors subdividing into an ensemble of small regions. Most regions have dimensions on length scales of nanometers, with atoms that are distinguishable from atoms in neighboring regions, thereby avoiding the need for macroscopic quantum symmetry for non-interacting point-like particles that may be meters apart. Atoms become indistinguishable only if they occupy the same region, where they are close enough to collide, or at least to have overlapping wavefunctions, consistent with the usual criterion for the onset of quantum behavior. If two macroscopic ensembles of these regions are combined, they meet the requirements for entropy of mixing, providing a novel solution to Gibbs' paradox by including finite-size effects in the thermodynamics.

Other results come from models based on Ising "spins" (binary degrees of freedom), which are solved analytically in 1-D. One example (**III E**) is a set of non-interacting Ising-like spins, with a total entropy that is increased by subdividing into separate spins, similar to the ideal gas. Another example (**III C**) is a 1-D chain of interacting Ising spins in contact with a bath of spins, which increases its entropy by forming a nanocanonical ensemble comprised of regions with a thermal distribution of all possible sizes. The average size is limited to two spins at high temperatures, diverging to large systems only as the temperature goes to zero. Equivalent results (**III D**) are found for a three-state model in the canonical ensemble with a large fixed size.

In summary, nanothermodynamics provides a systematic way to calculate contributions to energy and entropy across all length scales, providing new insight into the thermal behavior of old models. A general result is that the correct ensemble is needed to describe the thermal equilibrium of macroscopic systems, regardless of their size. Furthermore, distinct Hamiltonians can yield equivalent behavior, emphasizing the importance of including all contributions to energy that emerge from thermodynamics, not just those from microscopic interactions. Another result is a novel solution to Gibbs' paradox that does not require quantum symmetry of semi-classical ideal gas particles that may be meters apart. Further insight into the thermal behavior of other models may come from similarly strict adherence to the laws of thermodynamics across all length scales.

## V.  Acknowledgements

We have benefited from computer simulations done by Tyler Altamirano. We thank Sumiyoshi Abe for his careful reading of the manuscript, and his insightful comments.